\documentclass[conference]{IEEEtran}
\IEEEoverridecommandlockouts
\usepackage{cite}
\usepackage{amsmath,amssymb,amsfonts}
\usepackage{algorithmic}
\usepackage{graphicx}
\usepackage{textcomp}
\usepackage{xcolor}
\usepackage{caption}
\usepackage{multirow}
\usepackage{arydshln}
\def\BibTeX{{\rm B\kern-.05em{\sc i\kern-.025em b}\kern-.08em
    T\kern-.1667em\lower.7ex\hbox{E}\kern-.125emX}}
\begin{document}

\title{Dynamic Link Prediction Using Graph Representation Learning with Enhanced Structure and Temporal Information
}


\author{\IEEEauthorblockN{Chaokai Wu, Yansong Wang, Tao Jia\IEEEauthorrefmark{2}}
	\IEEEauthorblockA{
    \textit{College of Computer and Information Science} \\
	\textit{Southwest University}\\
	\textit{Chongqing, 400715, PR China} \\
	$[$chaokai0715,yansong0682,tjia$]$@gmail.com,@email.swu.edu.cn,@swu.edu.cn \\
	\IEEEauthorrefmark{2}Tao Jia is the corresponding author
	}
}

\maketitle

\begin{abstract}
	The links in many real networks are evolving with time. The task of dynamic link prediction is to use past connection histories to infer links of the network at a future time. How to effectively learn the temporal and structural pattern of the network dynamics is the key. In this paper, we propose a graph representation learning model based on enhanced structure and temporal information (GRL\_EnSAT). For structural information, we exploit a combination of a graph attention network (GAT) and a self-attention network to capture structural neighborhood. For temporal dynamics, we use a masked self-attention network to capture the dynamics in the link evolution. In this way, GRL\_EnSAT not only learns low-dimensional embedding vectors but also preserves the nonlinear dynamic feature of the evolving network. GRL\_EnSAT is evaluated on four real datasets, in which GRL\_EnSAT outperforms most advanced baselines. Benefiting from the dynamic self-attention mechanism, GRL\_EnSAT yields better performance than approaches based on recursive graph evolution modeling.
\end{abstract}

\begin{IEEEkeywords}
    Link prediction, Dynamic graph,
	Representation learning, Self-attention
\end{IEEEkeywords}

\section{Introduction}

Graphs are ubiquitous data structures that model pairwise interactions between entities \cite{b1}. Various complex systems can be represented as graphs, such as protein-protein interaction and human social behaviors, etc\cite{b2}. Most systems in the real world evolve over time, in which nodes and connections may disappear and recover. Such a dynamic system can be modeled as the dynamic graph \cite{b3}. 

As links of a network represent interactions between different entities, predicting future links is of great importance in the analysis of dynamic graphs. Dynamic link prediction has applications in many real-world problems, such as the discovery of new interactions between proteins \cite{a1}, the recommender system in social media and online shopping \cite{a2}, and more \cite{b2}. Following the success of applying neural networks for link prediction in static graphs \cite{b5,b6,b7}, researchers start to make use of the neural network for dynamic link prediction. By finding the representation of nodes, the link prediction problem is converted to the nearest neighbor search problem in the embedding space \cite{b8}. For dynamics graphs, the node representation relies on not only the structural properties but also the temporal patterns underlying the evolution of networks. Therefore, how to effectively extract and learn the temporal pattern is essential for the accurate prediction of future links \cite{b9}.

Traditionally, the learning of temporal patterns follows a recency assumption: old information is less important than recent information. This is implemented in the learning process by assigning weights that diminish with increasing duration \cite{b13}. Nevertheless, given a variety of dynamic systems, the recency assumption may not always hold. For example, when the network evolves periodically, information with a long history is more important to infer the temporal pattern. For this reason, we start to consider adding more flexibility in weighting historical data. In addition, the learning of structural features relies on a balance between local and global structures \cite{b23}. This is usually achieved by tuning the hyper-parameter \cite{a3}. But if the model can learn valid local and global information more intelligently, the performance can be further improved.

Motivated by these ideas, we explore and propose a new model GRL\_EnSAT for dynamic link prediction. GRL\_EnSAT utilizes the self-attention mechanism \cite{b18} that provides a flexible way to adjust the significance of historical and recent temporal information, as well as local and global structure information. The model can be roughly divided into structural learning and temporal learning block. In the structural block, the Graph Attention Network (GAT)\cite{b20} with a multi-head mechanism is used to learn structural information based on direct neighbors. Then an additional self-attention operation of distinguishing GAT independent heads is performed, letting the node obtain information about structures other than valid local neighbors from the remaining nodes. The structural representations of network snapshots at different times are sent to the temporal block, which combines these network snapshots using the mask self-attention network. GRL\_EnSAT is evaluated on four real networks, which demonstrates improved performance and good stability compared with baselines. We also perform ablation studies to quantitatively compare the contribution of structural and temporal blocks in GRL\_EnSAT. In general, GRL\_EnSAT benefits most from the application of a multi-head mechanism in temporal pattern learning.

\addtolength{\topmargin}{0.025in}

\section{Related Work}

\subsection{Non-deep-learning based Methods}
Matrix factorization is an effective tool for data processing, which includes Eigen decomposition, Singular Value Decomposition, etc. These techniques are widely applied in static or dynamic link prediction. Raymond et al. \cite{b26} generalize matrix factorization to the task of dynamic link prediction. The non-negative matrix decomposition (NMF), which decomposes the original matrix into two matrices with no negative elements, is also applied \cite{b27}. Ma et al. \cite{b13} propose an effective decomposition strategy for NMF to minimize the distance between two connected nodes in hidden space, which yields improved prediction accuracy. However, matrix factorization only considers structural information. Other useful features such as node attributes and temporal dynamics are overlooked. In addition, matrix factorization is often time-consuming, which limits its application. Therefore, feature-based models have received attention from scholars. Ran et al. \cite{b45} design a new similarity index by combining the shortest path characteristics of dynamic networks and the second-order neighborhood information. Ahmed et al. \cite{a10} introduce a sampling technique for similarity computation.  Wu et al.\cite{b46} introduce an aggregation mechanism to organize the most significant historical neighbors’ information and adaptively obtain the significance of node pairs.

\subsection{Deep Learning based Methods}
Inspired by the successes of deep learning in other fields, researchers started to apply deep neural networks in link prediction, with the aim to better combine both temporal and structural information \cite{CasSeqGCN, CCasGNN}. 
Nguyen et al. \cite{b31} redesign the random wandering strategy by adding constraints to the wandering process, such as considering temporal dependence and searching the obedient time, to learn more informative temporal embeddings. Hisano et al. \cite{b32} design the loss function that considers the supervised loss of past dynamics and the unsupervised loss of the current neighborhood context. Li et al. \cite{b15} propose a dynamic network embedding method based on the autoencoder, which uses historical information obtained from snapshots of the network with past timestamps. Similarly, Chen et al. \cite{b33} introduce an autoencoder with a recurrent neural network to dynamic link prediction. This model suits the networks of different scales with fine-tuned structures and prevents over-fitting through a regularization term. Lei et al. \cite{b34}  leverage the generative adversarial network to generate the next weighted network snapshot, which effectively solves the sparsity and the wide-value-range problem of edge weights in real-life dynamic networks. Hao et al. \cite{a7} consider the network dynamics from the node vector evolution sequence, which is modeled by a recurrent neural network. Since hyperbolic space has the better exponential capacity and hierarchy consciousness than Euclidean space, Yang et al. \cite{a8} map the structural and temporal information of the network to hyperbolic space for learning. For heterogeneous information networks, Zhan et al. \cite{b44} propose a balanced random walk strategy, which redesigns the traditional random walk with a sliding window into several short walks to avoid self-cycling and unbalanced node type problems.



\section{PRELIMINARY }

\begin{figure*}[htbp]
	\centering
	\includegraphics[width=1\linewidth]{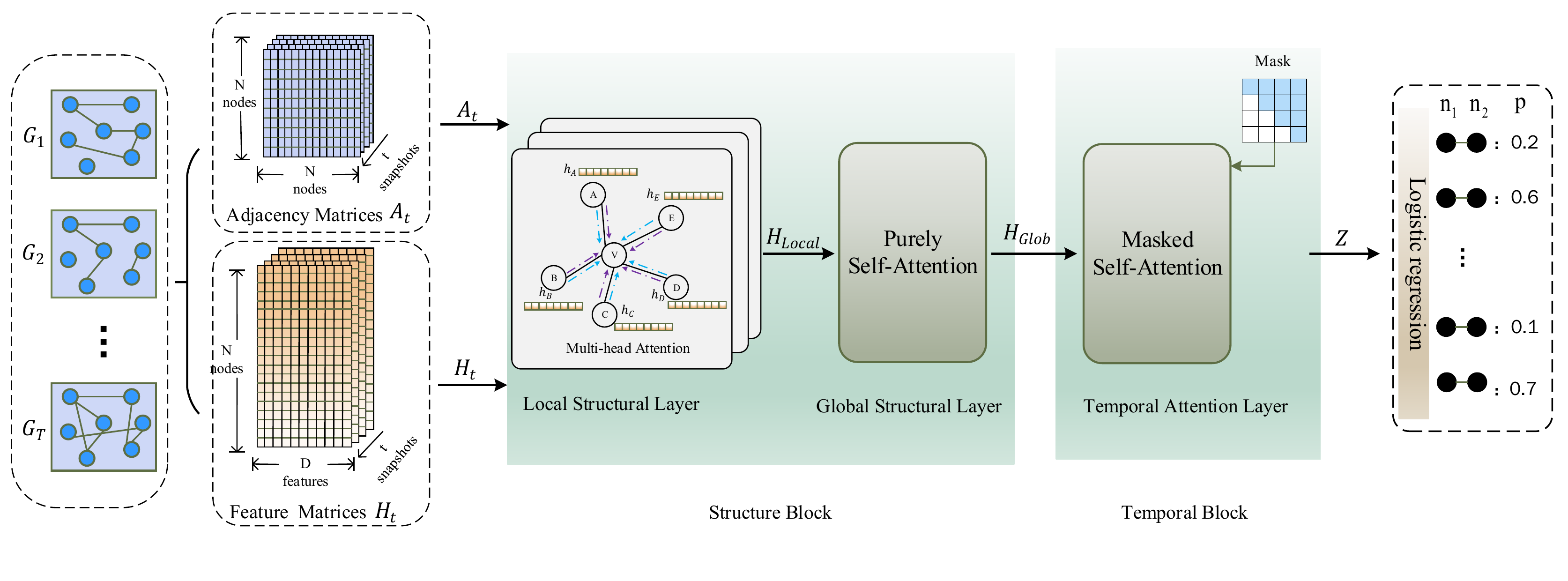}
	\captionsetup{font={small}}
	\caption{The overall structure of GRL\_EnSAT, decoupling dynamic graph information into two independent dimensions of `structural neighborhood', `temporal dynamics' for learning. }
	\label{fig:my_label1}
\end{figure*}

\subsection{Problem Definition}

\textit{\textbf{Definition 1} (Dynamic Networks)}: Consider a series of graphs (networks) $\left\{G_{1}, \ldots, G_{T}\right\}$, where $G_{k}=\left(\mathrm{V},\ \mathrm{E}_{k}\right), k \in [1,T],$ denotes the $k^{th}$ snapshot of the dynamic network. $V$ is the set of vertices of increasing number, suppose the total number of last moments is $N$ and $\mathrm{E}_{k} \subseteq \mathrm{V} \times \mathrm{V}$ is the set of links existed within the time window $\left[\mathrm{t}_{k-1}, \mathrm{t}_{k}\right]$. The adjacency matrix of the snapshot $G_{k}$ is denoted by $\mathrm{A}_{k}$, whose elements $\mathrm{a}_{k ; i, j}=1,i,j \in [1,N]$, if there is a link between $v_{i}$ and $v_{j}$, and otherwise $\mathrm{a}_{k ; i, j}=0$.

\textit{\textbf{Definition 2} (Dynamic Link Prediction)}: Given the matrix $\left\{A_{1}, A_{2}, \ldots, A_{T}\right\}$ of the past \textit{T} time snapshots, the goal of the dynamic link prediction is to predict the topology of the network at the next moment $T+1$. It can be formulated as
\begin{equation}
	\tilde{\mathrm{A}}_{\mathrm{T}+1}=f\left(A_{1}, A_{2}, \ldots, A_{T}\right),
\end{equation}
where $f(\cdot)$ is the dynamic link prediction model, and $\tilde{\mathrm{A}}_{\mathrm{T}+1}$ represents the predicted outcome.

\section{PROPOSED MODEL}
The structure of GRL\_EnSAT is shown in Fig. 1. The input of the model is the set of $T$ graph snapshots, and the output is representation vectors of all nodes at the next time step. In this section, we explain the modules of the model according to the order of data processing.

\addtolength{\topmargin}{0.025in}
\subsection{GRL\_EnSAT Architecture}
\textbf{\textit{Local structural layer.}} 
In this layer, the local structure information of network nodes is obtained by calculating the weights of direct neighbors. The operation is
\begin{equation}
	\begin{split}
		\overrightarrow{ \boldsymbol{h}_{v\ }^{\prime}}\!=\!\sigma&\left(\sum_{u \in N_{v}} \!\alpha_{v u} \boldsymbol{W} \overrightarrow{\boldsymbol{h}_{u}}\right),\  \alpha_{v u}=\frac{\exp \left(e_{v u}\right)}{\sum_{k \in N_{v}} \exp \left(e_{v k}\right)},\!\\
		e_{v u}=&\sigma\left(A_{v u} \cdot \boldsymbol{a}^\intercal\left[\boldsymbol{W}\cdot \overrightarrow{\boldsymbol{h}_{v}} \| \boldsymbol{W}\cdot \overrightarrow{\boldsymbol{h}_{u}}\right]\right), \forall(v, u) \in \varepsilon .
	\end{split}
\end{equation}

In the above calculations, $N_{v}$ is the set of direct neighbors of node $v$ in a graph snapshot $G$, $ \boldsymbol{W} \in \mathbb{R}^{D^\prime \times D}$ is a shared weight transformation applied to each node in the graph, $\overrightarrow{\boldsymbol{h}_{v}} \in \mathbb{R}^{D}$ is the initial vector of nodes $v$, $\boldsymbol{a}$ is a weight vector for the parameterized attention function, $.^\intercal$ denotes the transpose, $||$ is the concatenation operation and $A_{v u}$ is the weight of link $\left(v, u\right)$ in the current snapshot $G$.

For the stability of local structure information, we adopt the independent multi-head mechanism \cite{b20}. Through independent multiple operations, we obtain
\begin{equation}
	\boldsymbol{H}_{Local}(v)\!=\!\boldsymbol{\operatorname{Concat}}\!\left({\overrightarrow{\boldsymbol{h}_{v\ }^{\prime}}}^{1}\!,{\overrightarrow{\boldsymbol{h}_{v\ }^{\prime}}}^{2}\!, \ldots, {\overrightarrow{\boldsymbol{h}_{v\ }^{\prime}}}^{H_s}\right)\!\in\! \mathbb{R}^{D^\prime},
\end{equation}
where $H_s$ is the number of attention heads, and ${\overrightarrow{\boldsymbol{h}_{v\ }^{\prime}}}^{k}$ represents the representation of node $v$ learned from the $k^{th}$ independent head, the $Concat$ we choose is the operation of the connection. After repeating the above operations for all $T$ graph snapshots, we obtain a set of network representation matrix $\left\{\boldsymbol{H}_{Local}^{1}, \boldsymbol{H}_{Local}^{2}, \ldots, \boldsymbol{H}_{Local}^{T}\right\}, \boldsymbol{H}_{Local}^{t} \in \mathbb{R}^{N \times D^\prime}$ .

\textbf{\textit{Global structure layer.}} 
In this layer, we use a pure self-attention network with an independent multi-head mechanism to obtain global structure information beyond the direct neighborhood. To compute the output representation of node $v$, the scaled dot product form of attention \cite{b18} is used. The function is defined as
\begin{equation}
	\begin{split}
	{\boldsymbol{H}_{Glob}^t}^{h_s}=\boldsymbol{\operatorname{softmax}}(\frac{\boldsymbol{Q}_{Glob}^t \cdot {\boldsymbol{K}_{Glob}^t}^\intercal}{\sqrt{F^{\prime}}})\cdot \boldsymbol{V}_{Glob}^t \ ,
	\end{split}
\end{equation}
where the query ($\boldsymbol{Q}_{Glob}$), key ($\boldsymbol{K}_{Glob}$) and value ($\boldsymbol{V}_{Glob}$) are the results after transforming the representation of the input node through the trainable linear projection matrices $\boldsymbol{W}_{Q}^{\triangleright} \in \mathbb{R}^{D^\prime \times F^\prime}$, $\boldsymbol{W}_{K}^{\triangleright} \in \mathbb{R}^{D^\prime \times F^\prime}$ and $\boldsymbol{W}_{V}^{\triangleright} \in \mathbb{R}^{D^\prime \times F^\prime}$, respectively. They are formulated as
\begin{equation}
    \begin{split}
	\boldsymbol{Q}_{Glob}^t=\boldsymbol{H}_{Local}^t \cdot \boldsymbol{W}_{q}^{\triangleright},\\ \quad \boldsymbol{K}_{Glob}^t=\boldsymbol{H}_{Local}^t \cdot \boldsymbol{W}_{k}^{\triangleright},\\  \quad \boldsymbol{V}_{Glob}^t=\boldsymbol{H}_{Local}^t \cdot \boldsymbol{W}_{v}^{\triangleright}.
    \end{split}
\end{equation}

Through independent multiple operations, we obtain 
\begin{equation}
    \begin{split}
	\boldsymbol{H}_{Glob}^{t}\!=\!\boldsymbol{\operatorname{Concat}}\left({{\boldsymbol{H}_{Glob}^{t}}}^{1}, {{\boldsymbol{H}_{Glob}^{t}}}^{2},\ldots, {{\boldsymbol{H}_{Glob}^{t}}}^{H_s}\right),	
	\end{split}
\end{equation}
where ${{\boldsymbol{H}_{Glob}^{t}}}^{k}$ is the representation generated by the $k^{th}$ independent header. After conducting above operations on  $T$ snapshots, we feed ${\boldsymbol{H}_{Glob}^{t}} \in \mathbb{R}^{N \times F^\prime}$ into next module to learn temporal patterns.

\textbf{\textit{Temporal attention layer.}} Before proceeding to formal learning, the inputs are first reorganized. The set of vectors for each node across $T$ time steps is combined into a matrix $\boldsymbol{R}_{v} \in \mathbb{R}^{T \times F^\prime}, \forall v \in V$. Then, we use the self-attention network modified by Eq. 4, i.e., the masked self-attention network for learning.  The specific calculation process is simplified as follows,
\begin{equation}
    \begin{split}
        \boldsymbol{\boldsymbol{Z}_{v}}^{h_s}\!=\!\boldsymbol{\operatorname{softmax}}\!\left(\!\frac{\boldsymbol{R}_{v} \boldsymbol{W}_{Q}^{\circ} \!\cdot\!\left(\boldsymbol{R}_{v} \boldsymbol{W}_{K}^{\circ}\right)^{\intercal}}{\sqrt{F}}\!+\!\boldsymbol{M}\!\right) \!\cdot\! \boldsymbol{R}_{v} \boldsymbol{W}_{V}^{\circ},
    \end{split}
\end{equation}
where $\boldsymbol{W}_{Q}^{\circ} \in \mathbb{R}^{F^\prime \times F}$, $\boldsymbol{W}_{K}^{\circ} \in \mathbb{R}^{F^\prime \times F}$ and $\boldsymbol{W}_{V}^{\circ} \in \mathbb{R}^{F^\prime \times F}$ are all trainable linear projection matrices. $\boldsymbol{M} \in \mathbb{R}^{T \times T}$ is a masked matrix with each entry $M_{i j}=\{-\infty, 0\}$ to enforce the autoregressive property. To encode the temporal order, $\boldsymbol{M}$ is defined as
\begin{equation}
     M_{i j}=\left\{\begin{array}{l} 0, i \leq j \\
            -\infty, \text { otherwise }
\end{array},\right.
 \end{equation}
when $M_{i j}=-\infty$, the softmax function makes the attention weight equal to zero, closing the attention for time steps $i$ to $j$. After the independent multi-head operation, we get
\begin{equation}
    \boldsymbol{Z}_{v}=\boldsymbol{\operatorname{Concat}}\left({{\boldsymbol{Z}_{v}}}^{1}, {{\boldsymbol{Z}_{v}}}^{2}, \ldots, {{\boldsymbol{Z}_{v}}}^{H_{s}}\right) \in \mathbb{R}^{T \times F},     
\end{equation}
where ${{\boldsymbol{Z}_{v}}}^{k}$ denotes the result of temporal evolutionary information learning by the $k^{th}$ independent head on node $v$. The results can be reorganized as $\boldsymbol{Z}_{v} = \left[\boldsymbol{z}_{v}^{1}, \boldsymbol{z}_{v}^{2}, . ., \boldsymbol{z}_{v}^{T}\right]^\intercal$, where $\boldsymbol{z}_{v}^{t}$ denotes the representation of node $v$ at time step $t$. After all the nodes have gone through the above operations and the resulting representation $\boldsymbol{Z}$ can be used for prediction tasks and model parameter tuning.
 
\textbf{\textit{Prediction layer.}}
 We chose a logistic regression model with the following equation,
\begin{equation}
    y=\sigma\left(\boldsymbol{w}^{\intercal}\boldsymbol{x}\right)=\frac{1}{1+e^{-\boldsymbol{w}^{\intercal} \boldsymbol{x}}}\quad ,
\end{equation}
where $\boldsymbol{w}$ denotes the trainable weight, the input $\boldsymbol{x}$ is the combination of representations between two nodes, i.e., $\boldsymbol{x}=\left[\boldsymbol{z}_{v}^{T} \| \boldsymbol{z}_{u}^{T}\right] \in \mathbb{R}^{2F}, v,u \in V$, and the output $y$ is the probability of linking the edges of two nodes.

\subsection{Loss Function}
A binary cross-entropy loss is used. We use the representation of node $v$ at time step $t$ ($\boldsymbol{z}_{v}^{t}$) to maintain the local proximity of $v$ around $t$ \cite{b10}.
\begin{equation}
	\begin{split}
		L=\sum_{t=1}^{T} \sum_{v \in V}\Bigg(\sum_{u \in N_{w a l k}^{\;t}(v)}-\log \bigg(\sigma\Big( <\boldsymbol{z}_{u}^{t}, \boldsymbol{z}_{v}^{t}>\Big)\bigg)\\-w_{n} \cdot \sum_{u^{'} \in P_{\;n}^{\;t}(v)} \log \bigg(1-\sigma(<\boldsymbol{z}_{u^{'}}^{t}, \boldsymbol{z}_{v}^{t}>)\bigg)\Bigg),
	\end{split}
\end{equation}
where $\sigma$ is the sigmoid function, $<.>$ denotes the inner product operation, $u \in N_{w a l k}^{\;t}(v)$ is the set of nodes that co-occur with node $v$ in snapshot t for a fixed length random walk, $ p_{\;n}^{\;t}$ is the negative sampling distribution for snapshot $G_t$, and the negative sampling rate $w_n$ is an adjustable hyperparameter.

\section{EXPERIMENTS}

\subsection{Baselines, Metrics And Datasets  }
We consider both static and dynamic methods for dynamic link prediction. The static models are Node2vec\cite{b5}, Struc2vec\cite{b41} and SDNE\cite{b7}. The dynamic models include dyAERNN\cite{b6}, E-LSTM-D\cite{b33} and DySAT\cite{b10}. E-LSTM-D is an end-to-end link prediction model, which is able to provide probability values for predicted links. Except for E-LSTM-D, all other methods are actually network embedding methods. For these methods, We input the obtained network representation to the same logistic regression model to calculate link probability values. The predicted outcomes by these models are evaluated and compared using AUC and MAP. The specific details of the dataset used are summarized in Table I.

\begin{table}[h!]
	\centering
	\setcounter{table}{0}
	\renewcommand{\arraystretch}{0.95}
	\caption{Summary statistics for the four datasets.}
	\captionsetup{font={small}}
	\resizebox{\linewidth}{!}{
		\begin{tabular}{c|ccc}
			\hline 
			\text { Dataset } & \text { Nodes } & \text { Links } & \text { Time steps } \\
			\hline 
			\text { Enron\cite{b38} } & 143 & 22,784 & 16 \\
			\text { Fb-forum\cite{b39} } & 899 & 33,720 & 11 \\
			\text { Dept\cite{b43} } & 986 & 332,334 & 12 \\
			\text { UCI\cite{b40}} & 1,809 & 56,459 & 13 \\
			\hline
	\end{tabular}}
	\label{table:my_table2}
\end{table}

\subsection{Experimental Setup}

The model embedding size $d$ is set to 128, the number of independent attention heads is set to 8, and all historical snapshots are used for model learning in our evaluations. We combine the snapshots into a network and offer access to the network data by building an aggregation graph up to time $t$, where the link weights are proportional to the cumulative weights at time $t$ for static algorithms that cannot handle temporal dependencies \cite{b31, b15}. For the average of ten separate runs, the final findings are shown.


\subsection{Results}

\begin{table*}[ht]
	\centering
	\setcounter{table}{1}
	\captionsetup{font={small}}
	\caption{Results of dynamic link prediction experimental task based on four datasets. The best-performing models in each dataset are highlighted by the bolded operation, while the italic operation indicates the second best.}
	\renewcommand{\arraystretch}{1.1}
	\setlength{\tabcolsep}{1 mm}{
		\begin{tabular}{clllllllllll}
			\hline
			\multirow{2}{*}{Performance} & \multicolumn{2}{c}{Enron} & & \multicolumn{2}{c}{Fb-forum} & & \multicolumn{2}{c}{Dept} & & \multicolumn{2}{c}{UCI} \\
			\cline{2-3} \cline{5-6} \cline{8-9} \cline{11-12}  & \multicolumn{1}{c}{AUC} & \multicolumn{1}{c}{MAP} & & \multicolumn{1}{c}{AUC} & \multicolumn{1}{c}{MAP} & & \multicolumn{1}{c}{AUC} & \multicolumn{1}{c}{MAP} & & \multicolumn{1}{c}{AUC} & \multicolumn{1}{c}{MAP} \\ 
			\hline
			Node2vec &88.50$\pm$1.46  &61.03$\pm$0.67  &  &80.50$\pm$2.35  &57.39$\pm$0.95  &  &\textit{89.78}$\pm$1.10  &60.89$\pm$0.47 & &73.15$\pm$3.54 &54.19$\pm$1.28 \\
			Struc2vec &68.27$\pm$1.22 &53.51$\pm$0.37 & &74.43$\pm$2.43 &55.08$\pm$0.88 & &80.40$\pm$1.30 &57.49$\pm$0.58 & &74.56$\pm$3.00 &55.93$\pm$1.33 \\
			SDNE &63.06$\pm$3.60 &54.80$\pm$0.82 & &64.47$\pm$2.95 &57.48$\pm$0.59 & &73.88$\pm$2.78 &58.75$\pm$0.53 & &67.15$\pm$4.08 &57.84$\pm$0.90 \\
			DyAERNN &\textit{92.11}$\pm$0.86 &\textit{61.95}$\pm$0.54 & &74.96$\pm$3.33 &57.09$\pm$0.62 & &89.35$\pm$0.83 &\textbf{62.27}$\pm$0.28 & &83.30$\pm$3.00 &60.47$\pm$0.40 \\
			E-LSTM-D &89.47$\pm$0.92 &61.80$\pm$0.44 & &74.63$\pm$1.80 &55.91$\pm$0.49 & &89.08$\pm$0.48 &60.62$\pm$0.20 & &63.18$\pm$5.23 &52.63$\pm$1.05  \\
			DySAT &89.54$\pm$0.85 &60.74$\pm$0.65 & &\textit{82.77}$\pm$3.38 &\textit{57.50}$\pm$1.26 & &87.98$\pm$0.34 &60.48$\pm$0.29 & &\textit{90.70}$\pm$1.04 &\textit{61.35}$\pm$0.61  \\
			\textbf{GRL\_EnSAT} &\textbf{92.44}$\pm$0.80 &\textbf{62.35}$\pm$0.29 & &\textbf{86.59}$\pm$0.71 &\textbf{58.68}$\pm$0.43 & &\textbf{90.55}$\pm$0.40 &\textit{61.55}$\pm$0.43 & &\textbf{93.31}$\pm$0.18 &\textbf{62.25}$\pm$0.16 \\
			\hline
	\end{tabular}}
	\label{table:my_table4}
	\captionsetup{font={small}}
\end{table*}
\textbf{Overall Evaluation for Link Prediction}.
Table II presents the experimental findings, which clearly demonstrate that GRL\_EnSAT outperforms other baselines in all networks. The model improves the AUC by up to nearly 4 percentage points and the MAP by up to 1.2 percentage points when compared to the best baseline algorithm.

While it is intuitively expected that the dynamic model should perform better in dynamic link prediction, the static model can outperform the dynamic model in some cases. For instance, Node2vec outperforms almost all dynamic models in the Dept dataset except for GRL\_EnSAT. By synchronizing deep search and breadth search to gather enough neighborhood data, Node2vec achieves exceptionally nice performance. Compared with Node2vec, GRL\_EnSAT utilizes more temporal information which consequently helps it generate a more accurate prediction. For dynamic models that consider temporal patterns, DyAERNN and E-LSTM-D still perform less accurately than GRL\_EnSAT. We suspect the following factors may play a role. First, DyAERNN uses numerous layers of recurrent neural networks to capture long-term dependencies, but this approach may be less capable of capturing structural information than our model. Second, E-LSTM-D is an end-to-end supervised learning model that creates loss functions using an adjacency matrix, but this may be less effective in sparse networks. Finally, while both DySAT and our GRL\_EnSAT employ attention mechanisms to acquire representations, GRL\_EnSAT improves over DySAT due to the utilization of masked self-attention layer that better learns the temporal pattern.

\textbf{Snapshot quantity experiment}. Because the prediction of dynamic links is based on past snapshots. We analyze the AUC and MAP at each time step to examine the impact of the number of snapshots on the experiment (Fig. 2).

\begin{figure}[hpt]
	\centering
	\includegraphics[width=0.95\linewidth]{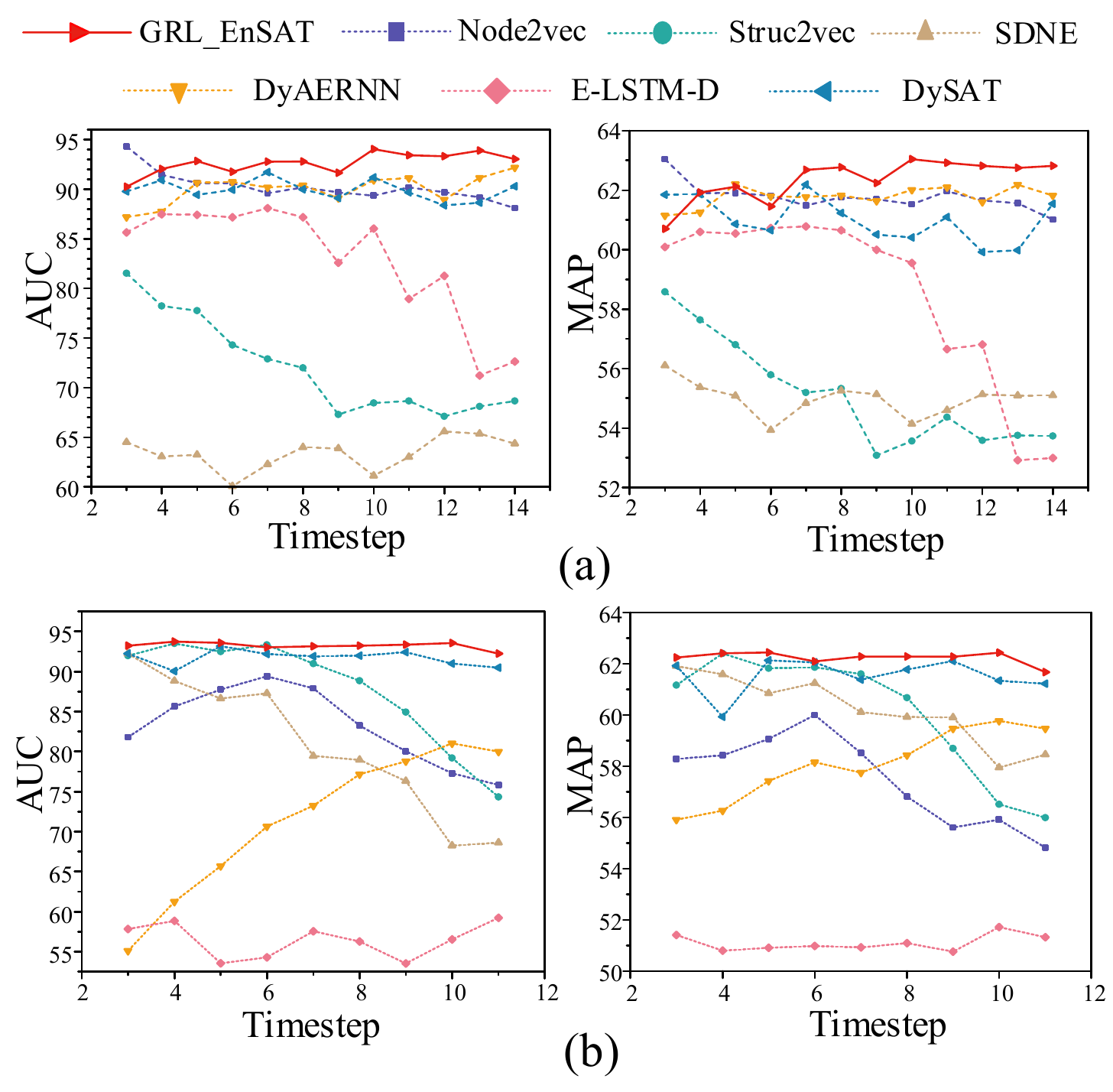}
	\captionsetup{font={small}}
	\caption{Experimental results on the quantity of snapshots. (a) Enron. (b) UCI.}
	\label{fig:my_label3}
\end{figure}

Fig. 2 demonstrates that GRL\_EnSAT almost outperforms all baselines under different sizes of historical training data, confirming the rubustness of the model. The outcome also suggests that more implicit qualities are captured by GRL\_EnSAT. In other words, GRL\_EnSAT is capable of capturing the network's fundamental evolutionary mechanism. For Enron network (Fig. 2a), Node2vec and DyAERNN perform well, indicating that the use of single feature, either the temporal evolution or the structural property, is sufficient for dynamic link prediction. But GRL\_EnSAT still performs better than others by fussing both structural and temporal information. For UCI network (Fig. 2b), the performance of the static models is noticeably less effective, which suggests that a static model cannot be used to analyze a dynamic scenario that is undergoing a long-term change. The performance of E-LSTM-D becomes worse when more historical data is used in training, implying its limit in adjusting parameters for temporal learning. In the contrary, DySAT is more stable. From the experiments on the two datasets, GRL\_EnSAT performs better than DySAT because it improves the learning of structure and evolution information.

\subsection{Analysis}
\textbf{Ablation Analysis.}
The local structure layer, the global structure layer, and the temporal layer are the three key components of our model. The ablation experiment is run on each part to investigate its contribution to the final outcome. For the sake of conciseness, the models are represented by the words `No Local', `No Global', and `No Temporal', whose performance is presented in Table III. The performance of `No Local' and `No Global' is worse than the original model, indicating that the two-layer structural self-attention is an effective operation. Compared with removing the structural self-attention layer, removing the temporal self-attention layer has a greater impact on the performance of the model. In Fb-forum, the model performance decreases by 20 and 3 percentage points in AUC and MAP indexes, respectively, which indicates the necessaries of learning temporal information in this dynamic network.



\begin{table}[htb]
	
	\centering
	\setcounter{table}{2}
	\captionsetup{font={small}}
	\caption{Ablation experiments performed on GRL\_EnSAT.}
	\renewcommand{\arraystretch}{1.1}
	\resizebox{\linewidth}{!}{
	\begin{tabular}{cllllllll}
			\hline
			\multirow{2}{*}{Performance} & \multicolumn{2}{c}{Fb-forum} & & \multicolumn{2}{c}{Dept} \\
			\cline{2-3} \cline{5-6}
			& \multicolumn{1}{c}{AUC} & \multicolumn{1}{c}{MAP} & & \multicolumn{1}{c}{AUC} & \multicolumn{1}{c}{MAP} \\ 
			\hline
			No Local &85.72$\pm$1.69 &58.23$\pm$1.05 & &90.04$\pm$0.15 &61.52$\pm$0.10  \\
			No Global &85.91$\pm$0.70 &58.37$\pm$0.39 & &90.38$\pm$0.55 &61.37$\pm$0.60 \\
			No Temporal & 64.49$\pm$0.92 &55.81$\pm$0.28 & &87.35$\pm$1.45 &60.15$\pm$0.64  \\
			original &\textbf{86.59}$\pm$0.71 &\textbf{58.68}$\pm$0.43 & &\textbf{90.55}$\pm$0.40 &\textbf{61.55}$\pm$0.43 \\
			\hline
	\end{tabular}}
	\label{table:my_table5}

	
\end{table}

\begin{figure}[ht]
	\centering
	\includegraphics[width=\linewidth]{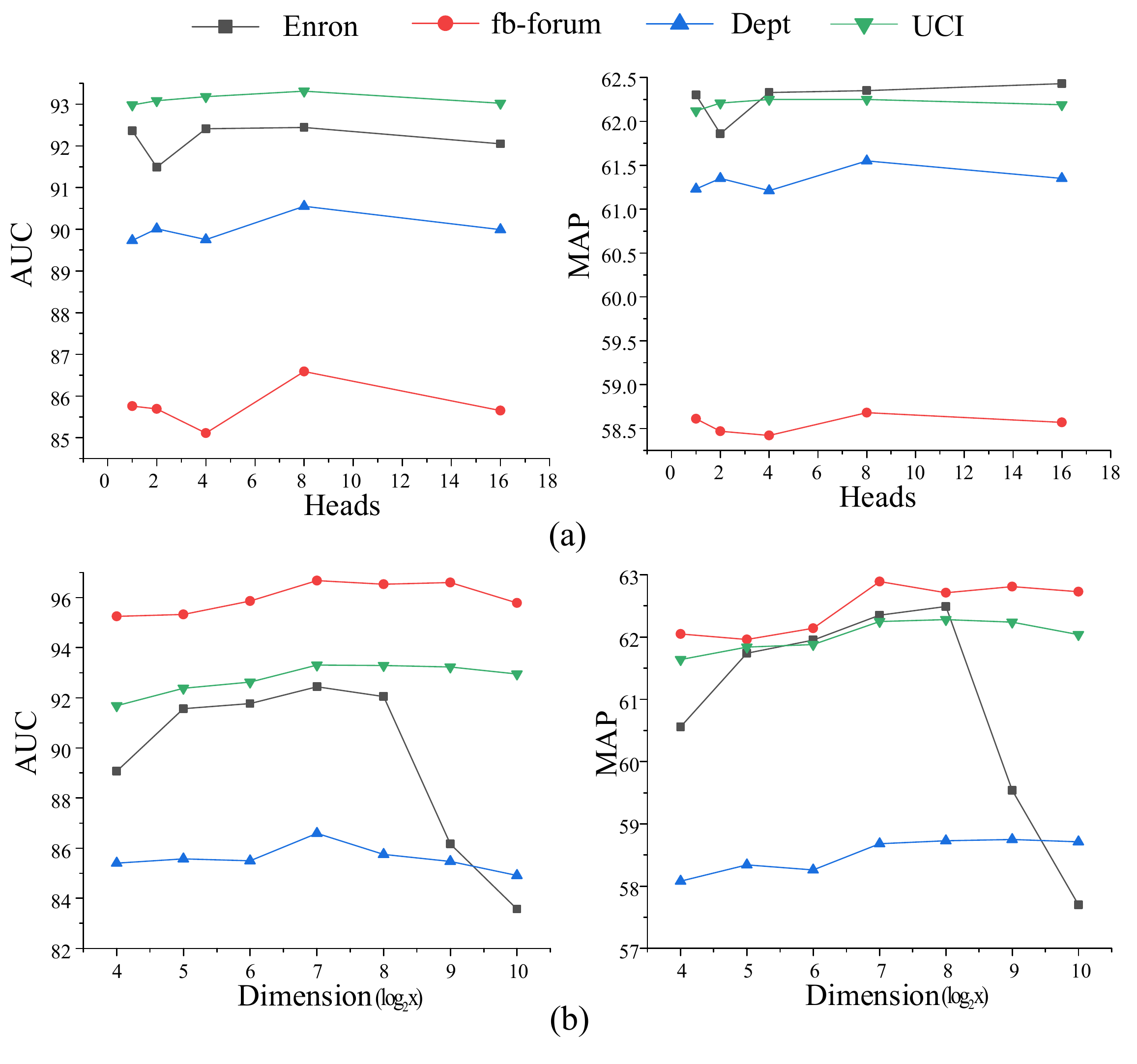}
	\captionsetup{font={small}}
	\caption{Experiments on the effect of model parameters on MAP and AUC values. (a) heads. (b) dimension($log_{2}x$).}
	\label{fig:my_label6}
\end{figure}

\textbf{Parameter Sensitivity Analysis.}
Since the model utilizes an independent multi-head mechanism, we conduct an experimental investigation of the number of multi-heads. We independently alter the GRL\_EnSAT's headcount while maintaining the other values. The ultimate outcomes are displayed in Fig. 3a. It is clear that when the number of multi-heads reaches 8, which appears to be sufficient to record the evolution of the graph in various possible ways, the curves perform at their best in terms of both AUC and MAP measures. This is substantial evidence that the separate multi-head setting mechanism is advantageous for GRL\_EnSAT.

We also compare the performance of GRL\_EnSAT at different embedding sizes (Fig. 3b). Consistent with our intuition, better accuracy and precision performance tends to occur at higher embedding dimensions. This is because low-dimensional vectors may have more information loss than higher-dimensional vectors. Our model shows consistently excellent performance in the embedding dimension of 128 $(2^7)$. This phenomenon demonstrates that GRL\_EnSAT has captured the essential features of the original network, and thus it can predict the unknown network well in highly compressed embeddings.


\section{CONCLUSION}
We present GRL\_EnSAT in this research as a way to capture network evolution and express it in response to graphs' dynamic character. GRL\_EnSAT picks up on the network's nonlinear characteristics as well as the temporal relationship between subsequent snapshots. Techniques for parameter inheritance are also utilized to keep the embedding stable and scalable. We do tests on four relevant datasets to test the model's validity. The outcomes demonstrate that it beats other baselines in terms of link prediction accuracy and precision. Even if the model is successful, additional data will inevitably be lost since the network is still represented discretely, with time as the primary deciding factor. Thus, continuous-time data formats with finer temporal granularity and less information loss will become more popular in the future.


\begin{thebibliography}{00}
	\addtolength{\textheight}{-1.5 cm} 
	\bibitem{b1} Pareja, A., Domeniconi, G., Chen, J., Ma, T., Suzumura, T., Kanezashi, H., Kaler, T., Schardl, T., and Leiserson, C., “EvolveGCN: Evolving Graph Convolutional Networks for Dynamic Graphs,” AAAI, vol. 34, no. 04, pp. 5363-5370, Apr. 2020.
	
	\bibitem{b2} P. Jiao, X. Guo, X. Jing, D. He, H. Wu, S. Pan, M. Gong, and W. Wang, “Temporal network embedding for link prediction via vae joint attention mechanism,” IEEE TNNLS, vol. 33, no. 12, pp. 7400-7413, Dec. 2022.
	
	\bibitem{b3} J. Sun, Y. Yang, N. N. Xiong, L. Dai, X. Peng and J. Luo, “Complex network construction of multivariate time series using information geometry,” IEEE Trans. Syst., Man, Cybern., Syst., vol. 49, no. 1, pp. 107–122, Jan. 2019.
	
	\bibitem{a1}  L. Hu, X. Wang, Y.-A. Huang, P. Hu, and Z.-H. You, “A survey on computational models for predicting protein-protein interactions,” Brief. Bioinf., Vol 22, no. 5, Sep. 2021.
	
	\bibitem{a2} S. Wang, L. Hu, Y. Wang, X. He, Q. Z. Sheng, M. A. Orgun, L. Cao, F. Ricci, and P. S. Yu, “Graph learning based recommender systems: A review,” IJCAI, pp. 4644–4652, 2021.


	\bibitem{b5} A. Grover and J. Leskovec, “Node2vec: Scalable feature learning for networks,” in Proc. 22nd ACM SIGKDD Int. Conf. Knowl. Discov. Data Mining, 2016, pp. 855–864.

	\bibitem{b6}  P. Goyal, S. R. Chhetri, and A. Canedo, “Dyngraph2vec: Capturing Network Dynamics using Dynamic Graph Representation Learning,” Knowl.-Based Syst., vol. 187, Jan. 2020, Art. no. 104816.

	\bibitem{b7} D. Wang, P. Cui, and W. Zhu, Structural deep network embedding,” in Proc. 22nd ACM SIGKDD Int. Conf. Knowl. Discovery Data Mining (KDD), 2016, pp. 1225–1234.

	\bibitem{b8} P. Goyal and E. Ferrara, “Graph embedding techniques, applications, and performance: A survey,” Knowl.-Based Syst., vol. 151, pp. 78–94, Jul. 2018.
			
	\bibitem{b9} J. Leskovec, J. Kleinberg, and C. Faloutsos, “Graph evolution: Densification and shrinking diameters,” ACM Trans. Knowl. Discovery Data, vol. 1, no. 1, 2007, Art. no. 2.
	
	\bibitem{b13}  X. Ma, P. Sun, and Y. Wang, “Graph regularized nonnegative matrix factorization for temporal link prediction in dynamic networks,” Physica A Statistical Mechanics \& Its Applications, vol. 496, 2018.
	
	\bibitem{b23} S. Cao, W. Lu, and Q. Xu, “Grarep: Learning graph representations with global structural information,” in Proc. 24th ACM Int. Conf. Inf. Knowl. Manage, 2015, pp. 891–900.
	
	\bibitem{a3} Q. Li, Z. Han, and X.-M. Wu, “Deeper insights into graph convolutional networks for semi-supervised learning,” in Proc. AAAI, 2018, pp. 1–8.
	
	
	\bibitem{b18} Vaswani, A., Shazeer, N., Parmar, N., Uszkoreit, J., Jones, L., Gomez, A. N., Kaiser, L., Polosukhin, I., “Attention is all you need,” in Proc. 31st Int. Conf. Neural Inf. Process. Syst., 2017, pp. 6000–6010.
	
	\bibitem{b20}  P. Veliˇckovic, G. Cucurull, A. Casanova, A. Romero, P. Lio, and Y. Bengio, “Graph attention networks,” in Proc. ICLR, 2018, p. 12.
	
	\bibitem{b26}  R. Raymond and H. Kashima, “Fast and scalable algorithms for semi-supervised link prediction on static and dynamic graphs,” in Proc. Joint Eur. Conf. Mach. Learn. Knowl. Discovery Databases, 2010, pp. 131–147.
	
	\bibitem{b27}  S. Gao, L. Denoyer, and P. Gallinari, “Temporal link prediction by integrating content and structure information,” in Proc. ACM 20th Int. Conf. Inf. Knowl. Manage., 2011, pp. 1169–1174.
	
	
	\bibitem{b45}  Y. J. Ran, S. Y. Liu, X. Y. Yu, K. K. Shang, and T. Jia, ‘‘Predicting future links with new nodes in temporal academic networks,’’ J. Phys., Complex., vol. 3, p. 11, Jan. 2022.
	
	\bibitem{a10} M. Ahmed, L. Chen, Y. Wang, B. Li, Y. Li, and W. Liu, ‘‘Sampling based algorithm for link prediction in temporal networks,’’ Inf. Sci., vol. 374, pp. 1–14, Dec. 2016.
	
	\bibitem{b46} Wu, J., Jia, T., Wang, Y., and Tao, L., ‘‘Significant Ties Graph Neural Networks for Continuous-Time Temporal Networks Modeling," arXiv preprint arXiv:2211.06590 (2022).

	
	\bibitem{CasSeqGCN} Wang, Y., Wang, X., Ran, Y., Michalski, R., and Jia, T., ‘‘CasSeqGCN: Combining network structure and temporal sequence to predict information cascades," Expert Systems with Applications, 206, Article 117693.
	
	\bibitem{CCasGNN} Wang, Y., Wang, X., and Jia, T., ‘‘Ccasgnn: Collaborative cascade prediction based on graph neural networks,’’ In 2022 IEEE 25th int. Conf. on Computer Supported Cooperative Work in Design (CSCWD) (pp. 810-815). IEEE.
	
	\bibitem{b31} G. H. Nguyen, J. B. Lee, R. A. Rossi, N. K. Ahmed, E. Koh, and S. Kim, “Continuous-time dynamic network embeddings,” in Proc. Companion Web Conf., 2018, pp. 969–976.
	
	\bibitem{b32}  R. Hisano, “Semi-supervised graph embedding approach to dynamic link prediction,” in International Workshop on Complex Networks, 2018, pp. 109–121.
	
	\bibitem{b15}  T. Li, J. Zhang, P. S. Yu, Y. Zhang, and Y. Yan, ‘‘Deep dynamic network embedding for link prediction,’’ IEEE Access, vol. 6, pp. 29219–29230, 2018.
	
	\bibitem{b33} Chen, J., Zhang, J., Xu, X., Fu, C., Zhang, D., Zhang, Q., and Xuan, Q., “E-LSTM-D: A deep learning framework for dynamic network link prediction,” IEEE Trans. Syst., Man, Cybern. Syst., vol. 51, no. 6, pp. 3699–3712, Jun. 2021.
	
	\bibitem{b34}  K. Lei, M. Qin, B. Bai, G. Zhang, and M. Yang, “GCN-GAN: A nonlinear temporal link prediction model for weighted dynamic networks,” in Proc. IEEE Int. Conf. Comput. Commun., 2019.
	
	\bibitem{a7}  X. Hao, T. Lian, and L. Wang, “Dynamic link prediction by integrating node vector evolution and local neighborhood representation,” in Proc. 43rd Int. ACM SIGIR Conf. Res. Develop. Inf. Retrieval, 2020, pp. 1717–1720.
	
	\bibitem{a8} M. Yang, M. Zhou, M. Kalander, Z. Huang, and I. King, “Discretetime temporal network embedding via implicit hierarchical learning in hyperbolic space,” in KDD, 2021, pp. 1975–1985.
	
	\bibitem{b44} Zhan, L., and Jia, T., ``Coarsas2hvec: Heterogeneous information network embedding with balanced network sampling," Entropy 24, no. 2: 276.
	
	
	\bibitem{b10} A. Sankar, Y. Wu, L. Gou, W. Zhang, and H. Yang, “Dysat: Deep neural representation learning on dynamic graphs via self-attention networks,” in Proc. ACM Int. Conf. Web Search Data Mining, 2020, pp. 519–527.
	
	
	\bibitem{b38} Klimt, B., and Yiming Y., ``Introducing the Enron corpus," CEAS. Vol. 45. 2004.
	
	\bibitem{b39} R. Rossi and N. Ahmed, “The network data repository with interactive graph analytics and visualization,” in Proc. AAAI, 2015, pp. 4292–4293.
 
	\bibitem{b43} A. Paranjape, A. R. Benson, and J. Leskovec, “Motifs in temporal networks,” in Proc. 10th ACM Int. Conf. Web Search Data Mining, 2017, pp. 601–610.
	
	\bibitem{b40} P. Panzarasa, T. Opsahl, and K. M. Carley, ‘‘Patterns and dynamics of users’ behavior and interaction: Network analysis of an online community,’’ J. Amer. Soc. Inf. Sci. Technol., vol. 60, no. 5, pp. 911–932, May 2009.
	
	\bibitem{b41} L. F. Ribeiro, P. H. Saverese, and D. R. Figueiredo, “Struc2vec: Learning node representations from structural identity,” in Proc. 17th ACM SIGKDD Int. Conf. Knowl. Discovery Data Mining, 2017, pp. 385–394.
	
	
	
	
\end{thebibliography}
\end{document}